# TOWARDS GENERALIZABLE LEARNING MODELS FOR EEG-BASED IDENTIFICATION OF PAIN PERCEPTION

*Mathis Rezzouk*[⋆,†], *Fabrice Gagnon*[⋆], *Alyson Champagne*[⋆],
Mathieu Roy[‡], Philippe Albouy[⋆], *Michel-Pierre Coll*[⋆], *Cem Subakan*[⋆,♭,†]

[⋆] Laval University, [†] Mila - Quebec Artificial Intelligence Institute,
[‡] McGill University, [♭] Concordia University

## ABSTRACT

EEG-based analysis of pain perception, enhanced by machine learning, reveals how the brain encodes pain by identifying neural patterns evoked by noxious stimulation. However, a major challenge that remains is the generalization of machine learning models across individuals, given the high cross-participant variability inherent to EEG signals and the limited focus on direct pain perception identification in current research. In this study, we systematically evaluate the performance of cross-participant generalization of a wide range of models, including traditional classifiers and deep neural classifiers for identifying the sensory modality of thermal pain and aversive auditory stimulation from EEG recordings. Using a novel dataset of EEG recordings from 108 participants, we benchmark model performance under both within- and cross-participant evaluation settings. Our findings show that traditional models suffered the largest drop from within- to cross-participant performance, while deep learning models proved more resilient, underscoring their potential for subject-invariant EEG decoding. Even though performance variability remained high, the strong results of the graph-based model highlight its potential to capture subject-invariant structure in EEG signals. On the other hand, we also share the preprocessed dataset used in this study, providing a standardized benchmark for evaluating future algorithms under the same generalization constraints. The code and dataset are available at: `https://github.com/Rinkachirikiari/eeg_pipeline_local.git`

*Index Terms*— Generalization, Cross-Participant, Machine Learning, Electroencephalography (EEG), Pain, Identification

## 1. INTRODUCTION

The advent of machine learning in the field of Brain-Computer Interfaces (BCI) has recently demonstrated significant potential in analyzing electroencephalographic (EEG) signals. These approaches have been successfully applied to domains such as detection of epileptiform discharges, sleep stage classification, and emotion recognition [1]. Although deep learning models have shown remarkable success in analyzing multivariate time series such as EEG signals, a major unresolved challenge remains their generalization across individuals. Unlike domains where data distributions are relatively stable across samples (e.g., computer vision or speech recognition)[2], EEG signals exhibit high cross-participant (cross-subject) variability due to differences in brain morphology, neural activity patterns, and electrode placements. This variability significantly impacts model robustness, limiting the applicability of EEG-based deep learning systems in real-world scenarios[1].

Machine learning approaches often perform well in within-participant [3] but fail when applied to new participants, highlighting the need for more adaptable learning strategies. Addressing this limitation could enable breakthroughs in clinical pain assessment, allowing pain monitoring without requiring patient-specific calibration. This study contributes to this open problem by systematically investigating the generalization performance of various machine learning models for EEG-based pain perception categorization on a new dataset regrouping 108 participants recorded in two sites. In this work, we applied both traditional classification and deep learning based classification approaches to assess their effectiveness in capturing relevant EEG patterns for distinguishing between painful thermal and aversive auditory stimuli. By comparing these models under within- and cross-participant evaluation, we aim to provide deeper insights into their generalization capabilities. Overall, our contributions can be summarized as follows:

- We introduce and publicly release the preprocessed data of a new EEG dataset on pain perception, offering a substantial and well-structured resource to advance machine learning research on pain classification. To the best of our knowledge, our dataset size is approximately five times larger than the median reported in prior EEG pain studies [4].

- We evaluate the generalization of various machine learning techniques to identify pain perception from

EEG signals in both within- and cross-participant paradigms, using a standardized preprocessing pipeline and consistent evaluation across 108 participants and multiple model families to ensure reproducibility. These results can serve as a baseline for further studies.

## 2. BACKGROUND

### 2.1. Pain Identification

The application of machine learning to pain research based on EEG has advanced significantly over the past decade, enabling complementary assessments of this inherently subjective experience. For instance, [5] proposed a method combining EEG signal processing with artificial neural networks to classify four levels of pain intensity, achieving 94.83% accuracy in within-participant experiments. Similarly, [4] conducted a comprehensive review highlighting the variety of AI methods from support vector machines to deep learning architectures employed to decode pain states, particularly in tasks involving pain intensity estimation and aversive stimulus detection. However, few studies have focused on distinguishing between different types of noxious or aversive stimulation based on EEG signals, rather than simply estimating pain intensity or relying on subjective ratings. In contrast, the ability to reliably distinguish between different perceptual states, such as aversive but non-painful stimulation versus genuinely painful experiences, remains largely unexplored. Moreover, [6] points out that many existing approaches lack external validation, raising concerns about their robustness and generalizability to unseen populations. These limitations highlight the need for further research on cross-participant generalization and on developing models capable of identifying pain perception itself, especially in real-world clinical contexts.

### 2.2. Generalizability

Ensuring cross-participant generalization in EEG classification remains a fundamental challenge due to the intrinsic variability of brain signals between individuals stemming from differences in head anatomy, electrode placement, neural dynamics, and noise levels [3]. One popular solution is to use transfer learning techniques, especially multi-source transfer learning, which have been employed to adapt knowledge from multiple participants to new ones. In [7], a method was proposed that leverages information from various source domains, including different participants and recording sessions, achieving strong cross-participant performance through adaptive feature extraction and transfer strategies. Interestingly, despite the prominence of deep learning, a substantial proportion of studies (27.4%) still rely on classical machine learning approaches [7], often achieving comparable results. For instance, [8] demonstrated that a self-organized graph-based method, which dynamically adapts to the input through adaptive filters, can perform competitively with domain adaptation strategies, suggesting feature customization at inference as a viable path for generalization.

Moreover, methods aiming at domain-invariant representation learning have explored architecture-level adaptations, such as domain generalization layers embedded in deep networks. In [9], for example, a flexible framework was proposed, combining such layers with loss functions designed to promote generalization, achieving promising results on both EEG and ECG data. While transfer learning strategies also show great potential in this context, particularly for leveraging information from multiple source domains, this study will focus on evaluating the generalization capabilities of various models across individuals through an extensive empirical analysis.

With regard to evaluation, the LOSOCV (*Leave-One-Subject-Out Cross-Validation*) protocol has become a widely adopted standard for evaluating the generalization performance of EEG classification models across participants, offering a reliable framework to uncover the limitations of conventional training schemes when applied to highly individual-specific EEG data [3].

## 3. DATASET

The EEG dataset that we are proposing with this study has been developed as part of a large-scale project aiming to collect data from 600 participants, with the goal of balancing generalizability and experimental feasibility. This sample size is approximately five times larger than the median reported in prior EEG pain studies [4]. For the present work, data from the 108 available participants were used (67 females, mean age: 25, SD: 6.3). Each participant took part in a single 3-hour experimental session conducted across two academic research sites and received monetary compensation. Participants were eligible if they were at least 18 years old, reported no chronic pain, no neurological or psychiatric disorders, and were not taking medications affecting attention or vigilance.

### 3.1. Stimulation Protocol

Two types of stimuli were applied during the experiment. Thermal painful stimulation was delivered using a TCS-II stimulator (QST Lab, France) or a MEDOC TSA-II stimulator (Medoc Ltd., Israel) on the non-dominant forearm, with temperatures calibrated individually for each participant based on their pain thresholds. Auditory aversive but non-painful stimulation consisted of high-pitched 1,000 Hz tones presented through headphones at variable intensities ranging from 50 to 90 dB. Each stimulation modality was presented twice: once during a passive task and once during an active task (see Figure 1). Each task block lasted approximately 8 minutes and was preceded by a 5-minute resting-state recording. The sequence of task presentation and stimulation modality was counterbalanced across participants to minimize order effects. For the purpose of this study, data from passive and active tasks were merged to focus exclusively on stimulus modality classification, independently of participants' behavioral responses.

### 3.2. EEG Acquisition

EEG was recorded using a 64-channel Acticap Slim cap (Brain Products, Germany), amplified by a NA300 (EGI, USA) or an ActiChamp (Brain Products, Germany) amplifier, and sampled at 1000 Hz. Electrodes were placed according to the international 10–20 system, with reference at FCz, ground at AFz, and impedances kept below 10 kΩ throughout data recording.

### 3.3. Open Science and Data Availability

The processed EEG data used in this study comprising recordings from 108 participants will be made publicly available (https://huggingface.co/datasets/MaRez10/EEG_Pain_Perception), and the dataset is organized according to the Brain Imaging Data Structure (BIDS) standard .

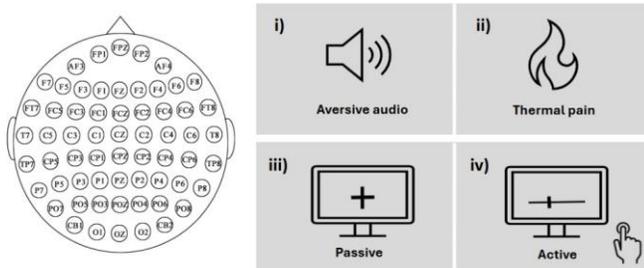

**Fig. 1**. Overview of the EEG data acquisition protocol. Participants underwent both aversive auditory (i) and painful thermal stimulation (ii), each lasting 8 minutes. For each type of stimulation, two task conditions were administered: a passive task (iii), during which no response was required, and an active task (iv), in which participants continuously rated the intensity of their perceived experience using response buttons, on a visual analog scale (VAS) with a scale of 0 (no sensation) to 200 (extremely unpleasant or painful).

## 4. METHOD

### 4.1. Preprocessing

EEG signals were band-pass filtered between 1 and 100 Hz to isolate relevant neural activity while removing slow drifts and high-frequency noise. A notch filter was applied at 60 Hz and its harmonics to eliminate line noise contamination. All signals were then downsampled to 250 Hz to reduce computational load while preserving the frequency components of interest. Continuous EEG recordings were segmented into fixed-length, non-overlapping epochs of 4 seconds, with each epoch labelled according to the type of stimulus presented during the corresponding time window. EEG epochs were normalized using z-score fitted on the training set and applying the transformation to both validation and test sets. This preprocessing pipeline was implemented using the MNE-Python library [10]. The chosen steps represent a targeted approach, prioritizing the preservation of genuine neural signals while addressing common data artifacts, acknowledging the value of minimal data manipulation [11]. The resulting preprocessed dataset of each participant was structured as a three-dimensional array of shape (500, 64, 1000), corresponding respectively to the number of samples, EEG channels, and time points.

### 4.2. Experimental Protocol

For dataset partitioning, data were split at the subject level into 70% training, 15% validation, and 15% test, aligning with the original splitting ratios used in the comparative deep models and following the optimal protocol proposed in [12]. In the participant-specific evaluation setting, each subject's data was independently divided using stratified 3-fold cross-validation to preserve label distribution across splits. For the cross-participant protocol, data from one participant was held out for testing, while the remaining individuals were used for training and validation. Within the training set, 30% of the unique participants were further set aside as a validation set to monitor generalization and guide early stopping. Each experiment was repeated with a fixed random seed across model families to ensure reproducibility.

### 4.3. Learning methods

We selected a subset of established machine learning approaches, both shallow and deep, that have shown strong performance in EEG-based classification tasks, offering a representative sample of state-of-the-art techniques [3], [6], [13]. Including shallow classifiers allows us to assess the performance of simpler, well-understood models that rely on handcrafted features and offer strong interpretability. This comparison provides a valuable baseline to evaluate the added benefits and limitations of deeper architectures.

#### 4.3.1. Shallow Classifiers

**1) CSP+SVM**[13] combines Common Spatial Patterns (CSP) with a Support Vector Machine classifier. CSP is a spatial filtering technique designed to maximize the variance between two classes, making it effective in discriminating between distinct brain states. Here, we apply CSP to extract discriminative spatial features between auditory and thermal pain con-ditions, followed by SVM for robust binary classification in a low-dimensional feature space.

**2) MDM**[14] (Minimum Distance to Mean) is a Riemannian geometry-based approach that operates on the covariance matrices of EEG trials. It assumes that class-specific covariance matrices lie on a Riemannian manifold and computes decisions based on geodesic distances to class means. This approach is particularly well-suited for noisy and non-stationary EEG data, capturing second-order statistical structure while being robust to outliers.

**3) TSLR**[13] (Tangent Space Logistic Regression) projects covariance matrices onto the tangent space of the Riemannian manifold, enabling the use of Euclidean classifiers such as logistic regression. This method balances geometric fidelity with computational simplicity, and is particularly relevant when dealing with cross-participant variability.

*4.3.2. Deep classifiers*

All deep learning models were implemented in PyTorch and trained on an NVIDIA Tesla P100 GPU. Each model was trained for up to 300 epochs using the Adam optimizer with a learning rate of 1e-3 (reduced on plateau) and a batch size of 32. Early stopping was applied with a patience of 50 epochs based on validation accuracy.

**4) Deep4Net**[2] is a deep convolutional neural network composed of a temporal convolution block, a spatial filter block, two additional convolutional layers, and a fully connected neural network. It is capable of learning hierarchical representations across time and space.

**5) ShallowFBCSPNet**[2] is a shallow variant that mimics CSP logic by learning frequency-specific spatial filters. It shares architectural components with Deep4Net but uses larger kernels, different activations, and pooling strategies.

**6) EEGNetv4**[15] is a compact model that uses depthwise and separable convolutions to efficiently learn temporal and spatial dependencies from EEG data.

**7) EEGConformer**[16] integrates convolutional blocks with transformer-based self-attention to combine local and global representations. Its structure includes temporal and spatial convolutions followed by self-attention layers, aiming to improve generalization to unseen participants.

**8) GGN**[17] (Gaussian Graph Network) leverages a learnable Gaussian graph generator to infer dynamic, sample-specific connectivity between electrodes by combining two sources of information: the physical distance between electrodes based on their montage positions, and the Pearson correlation between channels during each trial. These structural and functional topologies are merged into a single connectivity matrix by averaging them equally. The resulting graph enables the network to more effectively learn spatio-temporal patterns in EEG data, adapting to each participant and trial.

## 5. RESULTS

As shown in Table 1, shallow classifiers achieve strong performance in the within-participant setting, where training and test data share similar distributions. TSLR stands out with near-perfect accuracy, while CSP+SVM and MDM also perform well, highlighting the effectiveness of handcrafted spatial features and geometric representations when individual variability is controlled. However, in the cross-participant setting, performance drops sharply, exposing their limited ability to generalize across individuals.

| Model | Within | Cross |
| --- | --- | --- |
| CSP + SVM | 0.8674 ± 0.115 | 0.4734 ± 0.164 |
| MDM | 0.9161 ± 0.086 | 0.54261 ± 0.102 |
| TSLR | 0.9864 ± 0.023 | 0.57004 ± 0.091 |

**Table 1**. Comparison of test accuracy (± standard deviation) across shallow classifiers on all participants.

In Table 2, ShallowFBCSPNet and EEGConformer achieved notably good accuracy scores in the within-participant setup, while Deep4Net, EEGNetv4, and GGN showed signs of overfitting, based on their learning curves. In the cross-participant evaluation, Deep4Net obtained the highest accuracy, closely followed by GGN. Both models exhibited considerable variability across participants, suggesting that while they generalize better on average, their performance remains sensitive to inter-individual differences.

| Model | Within | Cross | $\theta$ |
| --- | --- | --- | --- |
| ShallowFBCSPNet | 0.8917 ± 0.123 | 0.8512 ± 0.205 | 108K |
| Deep4Net | 0.5959 ± 0.147 | 0.8606 ± 0.207 | 306K |
| EEGNetv4 | 0.5845 ± 0.165 | 0.6500 ± 0.120 | 3.1K |
| EEGConformer | 0.8363 ± 0.141 | 0.6600 ± 0.205 | 856K |
| GGN | 0.5193 ± 0.050 | 0.8523 ± 0.193 | 1.5M |

**Table 2**. Comparison of average test accuracy (± standard deviation) deep classifiers on all participants along with the number of trainable parameters ($\theta$).

## 6. DISCUSSION

### 6.1. Results interpretation

Classical machine learning models achieved strong performance in the within-participant setting but showed a marked decline in cross-participant evaluations, reflecting their sensitivity to inter-individual variability. Among these approaches, TSLR consistently outperformed other classical models across both settings. This robustness can be attributed to the tangent space projection allowing conventional classifiers to operate more effectively, echoing prior findings in [18]. In contrast, deep learning pipelines were evaluated using their original hyperparameters without dataset-specific optimization nor data augmentation, as tuning complex architectures like kernel sizes or activation functions was beyond the scope of this benchmark. These methodological choices were intentional, reflecting the aim to assess models in their canonical, non-tuned implementation rather than to maximize performance through intensive fine-tuning. Deep learning models, despite their strong average performance in cross-participant evaluation, exhibited substantial variability across individuals (see Figure 2) that might come from the low signal-to-noise ratio of EEG signals and the risk of overfitting non-informative patterns when training flexible ConvNet architectures without extensive regularization [2] as further illustrated by post-hoc statistical analyses and learning curves available in the repository appendix [1]. Interestingly, ShallowFBCSPNet, with its much shallower architecture, achieved near-comparable cross-participant performance to Deep4Net, suggesting that simply increasing model depth is not the most effective strategy for generalization in EEG decoding. In contrast, GGN, which incorporates spatially adaptive graph modeling, delivered the second-best generalization performance, supporting the idea that modeling dynamic electrode connectivity provides a more robust pathway toward capturing subject-invariant features.

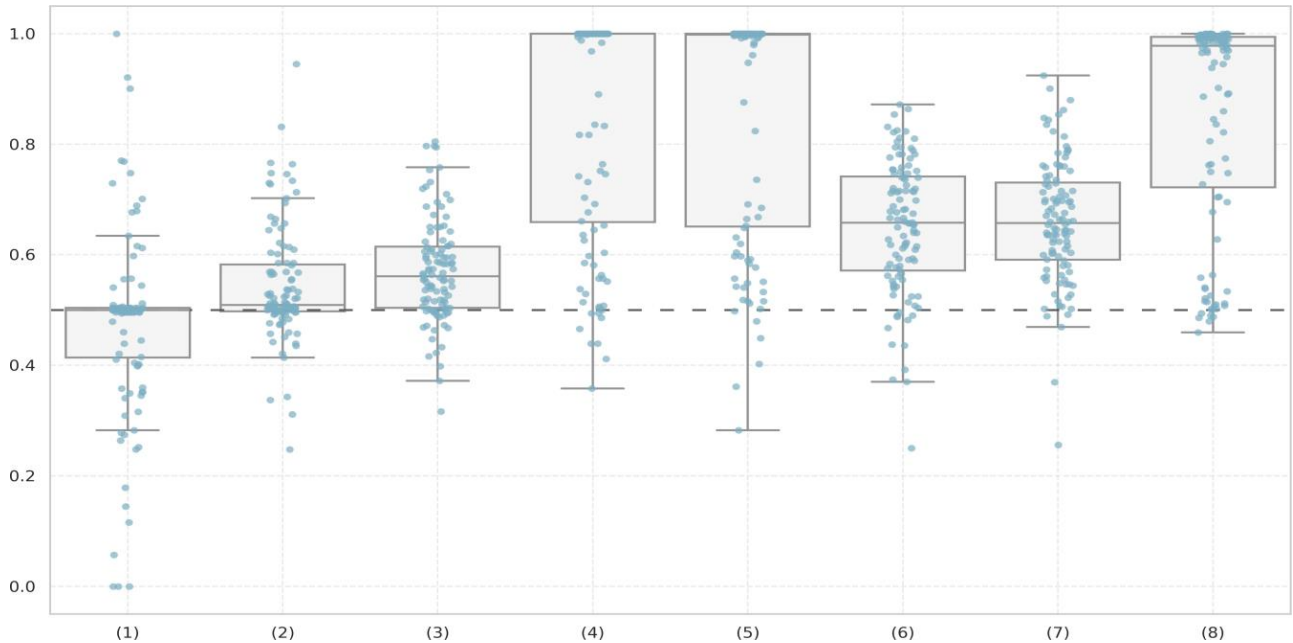

**Fig. 2**. Cross-participant distribution of test accuracies across individuals. The models are respectively : (1) CSP, (2) MDM, (3) TSLR, (4) ShallowFBCSPNet, (5) Deep4Net, (6) EEGNetv4, (7) EEGConformer, and (8) GGN, highlighting performance variability across models.

**6.2. Future Work**

While this study provides valuable insights into the generalization capabilities of machine learning models for EEG-based sensory state identification, several avenues remain open for future exploration like [19] which motivates the integration of self-supervised learning (SSL) methods to leverage unlabelled EEG data and enhance feature robustness across participants. A promising direction would be to extend graph-based models such as GGN by integrating self-supervised learning techniques. Coupling spatially adaptive graph architectures with unsupervised pretraining could promote the extraction of invariant and transferable EEG representations. As a next step, future work could focus on performing regression directly on participants' continuous pain ratings, aiming to predict different levels of perceived pain and move beyond binary classification of stimulation modalities. Finally, exploring model post-hoc techniques, such as spatial attention mapping or layer-wise relevance propagation, would offer important insights into how models associate neural patterns with pain perception and could increase trust in future clinical applications.

---

[1] Appendix available at https://github.com/Rinkachirikiari/eeg_pipeline_local/blob/main/appendix.md

## 7. CONCLUSION

This work presents a comprehensive evaluation of machine learning and deep learning models for the identification of perceived pain from EEG recordings, with a particular focus on cross-participant generalization. We share a preprocessed large-scale dataset of 108 participants exposed to two distinct stimulation modalities; we compared the performance of classical classifiers and modern neural architectures under both within- and cross-participant evaluation schemes. Our results confirm that while traditional models can achieve high accuracy in participant-specific settings, their performance deteriorates significantly when tested on unseen individuals. In contrast, deep learning models exhibited greater resilience to inter-individual variability, suggesting their potential for real-world applications where participant calibration is not feasible. This study establishes a benchmark for generalization in EEG-based pain identification. While offering new opportunities, the dataset also presents significant challenges, such as high inter-individual variability and limited signal-to-noise ratio, making it a valuable test bed for developing more generalizable and robust EEG decoding methods. We invite the community to further explore these challenges.